
\newif\ifOmitFigures


\ifOmitFigures\message{FIGURES WILL BE OMITTED}\else\message{FIGURES WILL
BE INCLUDED}\fi \ifOmitFigures\def\input#1#2#3#4{{}}\def\epsfverbosetrue{}
\def\special#1{\vskip 5 truecm}\def\epsfbox#1{\vskip 5 truecm}\fi
\newdimen\epsfxsize\newdimen\epsfysize



\def\Integers{\hbox{${\rm Z \kern 0.3ex \llap{Z}}$}}


%
\let\miguu=\footnote
\def\footnote#1#2{{$\,$\parindent=9pt\baselineskip=13pt%
\miguu{#1}{#2\vskip -5truept}}}
%

     \def\=>{\Rightarrow}
\def \==> {\Longrightarrow}
\def\dal{\displaystyle{{\hbox to 0pt{$\sqcup$\hss}}\sqcap}}

\def\to{\rightarrow}
\def\tilde{\widetilde}

\def\ideq{\equiv}
\def\Reals{{\rm I\!\rm R}}


\def\interior #1 {{ \buildrel\circ\over{{#1}} }} 
\def\interior #1 {  \buildrel\circ\over  #1}     
\def\Lie{\hbox{{\it\$}}}  
\def\lto { {\raise1pt\hbox{$<$}} \!\!\!\! {\lower4pt\hbox{$\sim$}} }
\def\gto { {\raise1pt\hbox{$>$}} \!\!\!\! {\lower4pt\hbox{$\sim$}} }



\input epsf
\epsfverbosetrue

\def\FigureNumberCaption #1#2#3 {\vbox{
\centerline{\vbox{\epsfbox{#1}}}
\leftskip=1.5truecm\rightskip=1.5truecm     
\singlespace
\noindent{\it Figure #2}. #3
\vskip .25in\leftskip=0truecm\rightskip=0truecm}
\sesquispace}

\def\singlespace{\baselineskip=12pt}
\def\sesquispace{\baselineskip=14pt plus 2pt minus 1 pt}
\raggedbottom
\magnification=\magstep1






\def\dtA{{d^2\!A}}  
\def\L{{\cal L}}
\def\d{\delta}
\def\X{{\cal X}}
\def\ptl{\partial}
\def\T{{\cal T}}
\def\W{{\cal W}}
\def\div{{\rm div}}
\def\tr{{\rm tr}}

\def\subsection #1 {\medskip\noindent{\it [ #1 ]}\par\nobreak\smallskip}
\def\section    #1 {\bigskip\noindent{\bf   #1  }\par\nobreak\smallskip}



\singlespace
\vskip -1 true cm
\rightline{SU--GP--95--7--3}
\rightline{gr-qc/9508002}
\bigskip
\sesquispace
\centerline{\bf Two Topics concerning Black Holes: Extremality of the}
\centerline{\bf Energy, Fractality of the Horizon\footnote{*}
 {To appear in
 Steve Fulling (ed.), Proceedings of the Conference on Heat Kernels and
 Quantum Gravity, held August, 1994, Winnipeg, Canada (U. of Texas). }}
\singlespace  
\bigskip
\centerline {\it Rafael D. Sorkin}
\medskip
\centerline
{\it Department of Physics, Syracuse University, Syracuse, NY 13244-1130}
\smallskip
\centerline { \it and }
\smallskip
\centerline {\it Instituto de Ciencias Nucleares, UNAM, A. Postal 70-543,
             D.F. 04510, Mexico.}
\smallskip
\centerline {\it internet: rdsorkin@mailbox.syr.edu}
\singlespace
\bigskip\leftskip=1.5truecm\rightskip=1.5truecm     
\centerline{\bf Abstract}
\medskip
\noindent
We treat two aspects of the physics of stationary black holes.  First we
prove that the proportionality, $d(energy) \propto d(area)$ for arbitrary
perturbations (``extended first law''), follows directly from  an
extremality theorem drawn from earlier work [1].  Second we
consider quantum fluctuations in the shape of the horizon, concluding
heuristically that they exhibit a fractal character, with order $\lambda$
fluctuations occurring on all scales $\lambda$ below $M^{1/3}$ in natural
units.
\bigskip\leftskip=0truecm\rightskip=0truecm 	    
\sesquispace                                 	    
\bigskip\medskip


   The theory of black hole thermodynamics is incomplete.  On one hand the
identification of black hole entropy with horizon area seems established by
a preponderance of direct and indirect evidence.  On the other hand we are
still in the dark about the physical variables whose ``states'' this entropy
counts. (For a recent review see [2].)

   The two main sections of this paper belong with the two ``hands'' just
mentioned.  The first provides a new proof of one of the main pieces of
evidence for the thermodynamical interpretation of black hole properties,
namely the ``first law'' (in its extended form which deals with arbitrary
variations, not just stationary ones).  It is essentially the text of my
talk at the conference.  The second main section presents some evidence for
a fractal structure of the horizon in the context of contributions to the
entropy from fluctuations in ambient quantum fields and fluctuations in the
shape of the horizon itself.  It reports on some ideas I discussed
informally with participants in the conference, especially Andrei Zelnikov
and Valeri Frolov.

\section{ I. Derivation of the ``First Law''}

The work I will describe in this section, done together with Madhavan
Varada\-rajan [3], grew out of our wish to understand what happens to
the theorem that stationarity implies extremality, when spacetime has a
boundary.  It has been known for a long time that for gravity, or any other
Lagrangian field theory, every solution of the field equations which has a
Killing vector also has a corresponding extremality property: the conserved
quantity associated to the Killing vector is unchanged by infinitesimal
perturbations of the fields.  Bernard Schutz and I had found a proof of
this which we liked [1] and we wondered at the time what would
happen if we applied it to a spacetime containing a black hole.  The main
message I want to leave you with today, is that what happens is that the
so-called first law of black hole thermodynamics emerges in a very direct
manner.

The derivation which results in this way is of interest mainly because of
its conceptual simplicity, but it also shows one new thing.  It shows that
the 3-surface $\Sigma$ on which the energy is evaluated can meet the black
hole horizon anywhere; it doesn't have to go through any special place like
a bifurcation submanifold\footnote{$^{1}$}
{as it does, for example, in reference [4].}.
I believe this is important, because the ability to push $\Sigma$ forward
along the horizon is crucial to understanding where the {\it second} law of
black hole thermodynamics comes from [5].

  The proof also makes  clear why the first variation of the energy gets
contributions only from the horizon itself, and it provides an explanation
(via the Raychaudhuri equation) of why it is specifically the change in
horizon {\it area} which governs the change in the energy.

The derivation also illustrates how integral
formulations of conservation laws can often be more convenient than
differential ones.  It takes place in 4D for Einstein gravity (with a
possible electro-magnetic field), but there is no reason it could not be
extended to higher dimensions, or to other lagrangians.  The proof is also
in such a form that it might help in understanding the behavior of the {\it
second} variation of the energy.  This variation is important in connection
with stability, but I don't have any new results to report on it.

Since a detailed account will soon be available [3], there is no
reason to try for completeness here.  Instead I will  present the
main steps of the analysis as simply as I can, in a manner which I hope
will be complementary to that of reference [3].  In the same
minimalist spirit I will mainly restrict myself to the case of pure
Einstein gravity and will set the electromagnetic field and black hole
rotation rate to zero.

\subsection{The Noether operator and the total energy}

Before we can get to the proof proper, we need the notion of Noether
operator and a technique I will call asymptotic patching.  The Noether
operator formalizes her explanation of how continuous symmetries imply
conservation laws.  For a first-order Action $S$ depending on dynamical
fields $Q$ and background fields $B$, and for a geometrical symmetry like
energy or angular momentum, the Noether operator is defined through the
identity,
$$
   \d_{(f,\xi)} S =
                   \oint_{\ptl\X} f \, \T^a_b\cdot\xi^b d\sigma_a
                 - \int_\X {\d \L \over \d Q} \, f \Lie_\xi Q \, dV
                                                            \eqno(1)
$$
Here the variation $\d_{(f,\xi)}$ is what might be called a ``partial
dragging'' of both the fields $Q$ and the region of integration $\X$
through the vectorfield $\xi$, specifically it drags $\ptl\X$ by an amount
$f\xi$ and it alters $Q$ by $\d_{(f,\xi)}Q = - f\Lie_\xi Q$.  If there are
no background fields $B$ present in the lagrangian $\L=\L[Q;B]$ --- or if
$\xi$ is a symmetry of those which are present --- then for $f\ideq 1$, the
variation $\d_\xi{S}$ evidently vanishes identically.

Now the total energy of a solution, evaluated on a surface $\Sigma$
which is the future boundary of a spacetime region $\X$, can be defined as
the value of $\d S(\X)$ when $\Sigma$ and all the fields on it are
translated in time in such a manner as to hold fixed the boundary
conditions at infinity [6].  Choosing $f$ and $\xi$ to implement
these requirements and assuming sufficiently rapid falloff of $Q$ at
infinity leads directly to the formula for the energy
$$
      - E = \int_\Sigma \T^a_b\cdot\xi^b \, d\sigma_a,  \eqno(2)
$$
where $\xi$ is any vectorfield which is a (future-directed)
time translation in a neighborhood of infinity (an exact Killing vector of
the flat background there.)

In a situation where the background either is absent or enters only as a
surface term in the Action, a compensating deformation by $-f\xi$ reduces
$\d{S}$, and therefore $E$, to a surface integral at spatial infinity.  In
virtue of (2), this has the formal consequence that
$\T^a_b\cdot\xi^b$ must take the form of a pure divergence
$\ptl_b(\W^{ab}\cdot\xi^b)$ plus a term which vanishes ``on shell''.
Specializing to gravity (and for brevity omitting indices and indications
of elements of surface/volume and of density-weight), we have specifically
$$
     T \cdot \xi = \div (W\cdot\xi) - G \xi,   \eqno(3)
$$
so that (2) reduces on shell to\footnote{$^{2}$}
{For angular momentum, everything would be the same, except that $\xi$
would be a rotation generator near spatial infinity, instead of a time
translation.  The explicit form of $W$ can be found in refs. [6],
[7] and [3].}
$$
      - E = \oint_\infty W \,  .          \eqno(4)
$$

\subsection{Asymptotic patching}

  In computing the energy, etcetera, of an asymptotically flat solution
$g_{ab}$ to the Einstein equation, we don't directly use the covariant
Action ${1\over{2}}\int{RdV}$.  Instead we do something else which can be
described in different ways.  Perhaps the best description is just that we
replace the metric $g$ by one which is strictly flat near spatial infinity.
In reality the very long range part of the metric representing an isolated
system like a black hole is meaningless in any case, since it cannot be
isolated from the fields of other objects which are invariably present.
Hence, there should be no distinction, in a physical sense, between
$g_{ab}$ and a metric which has been ``cut off'' at large radii by
``patching'' it to a flat metric.  The $S$ whose variation yields the
energy of the isolated system is best viewed, I believe, as nothing but the
covariant Action\footnote{$^{3}$}
{For many purposes, one also needs to add to this Action a surface term
 like $\tr{K}$ integrated over the initial and final spacelike boundaries;
 but here we can ignore any such addition since it does not contribute to
 the variation defining $E$.}
of this cut off field $\tilde g_{ab}$; and the technique for obtaining
$\tilde g_{ab}$ by gradually deforming the original metric to the flat one
as some radial parameter $r$ increases from $R$ to $2R$ is what I mean by
``asymptotic patching'' [1][3].

  Asymptotic patching can also be understood in a purely technical way in
relation to an integration by parts performed to render the Action finite.
For generic O($1/r$) falloff in the metric, the Ricci scalar $R$ will decay
only like $1/r^3$, which leads to a logarithmically divergent integral for
$S$.  By adding a suitable divergence to the integrand we eliminate from
$R$ the terms of the form $g\ptl\ptl{g}$, thereby improving its falloff to
$1/r^4$, while at the same time making $S$ first-order so that the above
definition of the Noether operator applies without modfication.  The
improved falloff suffices to render both $S$ and $E$
well-defined.\footnote{$^{4}$}%
{For angular momentum a strengthened asymptotic condition is needed
(``parity condition'').}
Having modified $S$ in this way, we can then perform the same patching to a
flat metric at infinity without producing any further change in the Action
or the energy (in the limit in which the patching radius $R$ recedes to
infinity) [1][3].  This second viewpoint is perhaps somewhat
more advantageous technically, but it requires the introduction of extra
background: a globally defined connection with respect to which one can
perform the integration by parts.

The upshot from either point of view is that we end up having to deal only
with metrics which are strictly flat near infinity.  This will free us from
having to worry about the effects of variations at infinity, leaving only
boundary terms at the horizon to contribute.  It also means, of course,
that we can no longer express the energy as the flux integral (4) taken
at true infinity, but (4) still holds if evaluated {\it just inside}
the patching radius, and the expression (2) in terms of a spatial
integral remains generally valid, under the assumption (which will always
be in force) that $\xi^a$ remains an exact Killing vector of the flat
asymptotic metric throughout the patching region.

Henceforth, we will consider only metrics which have been patched to become
strictly flat near $\infty$, and "solution" will always mean solution
patched to flat metric at large $r$.  In addition, we will consider only
vectorfields $\xi$ which preserve any background which has been introduced,
and which in particular are strict Killing vectors (of the flat metric)
near infinity.

\subsection{The extremum proof without reference to a horizon}

Setting $f=1$ in the defining identity (1) of the Noether operator, and
recalling that the left hand side then vanishes automatically, we obtain
the basic identity
$$
  \oint T\cdot\xi  = \int {\d L\over\d g} \Lie_\xi g ,     \eqno(5)
$$
for an arbitrary metric $g$ and vectorfield $\xi$.  (Here $\d L / \d g$, if
made explicit, would be $-G^{ab}$ of course.)

Now consider (Figure 1) a spacetime region $\X$ bounded to the past and
future by asymptotically flat surfaces $\Sigma_0$ and $\Sigma$, and let the
metric $g$ be {\it a stationary solution} to the Einstein equation.

\epsfxsize=3.6in
\FigureNumberCaption{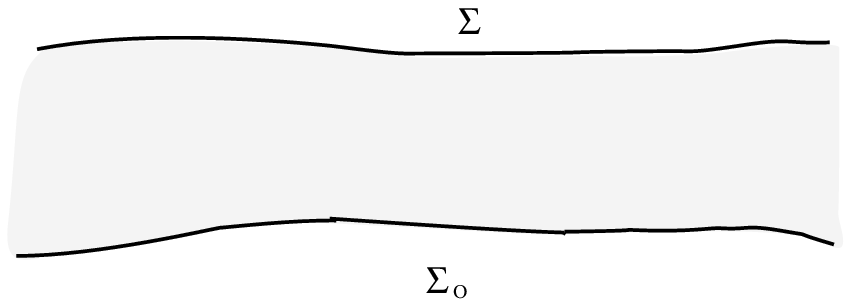}%
{1}%
{The spacetime region $\X$ involved in proving energy
extremality without reference to a horizon.  It is bounded to the future
and past by the surfaces $\Sigma$ and $\Sigma_0$.}

\noindent
If $g'$ is a nearby solution (not necessarily stationary) then it is easy to
cobble together a perturbation $\d g$ which vanishes in a neighborhood of
$\Sigma_0$ and for which $\d g = g'-g \;$ in a neighborhood of $\Sigma$.
Let us apply the identity (5) to this perturbation, in fact let
us consider the result of perturbing $g$ in (5) by an arbitrary
$\d g$.  On the RHS we have the product of two expressions which both
vanish for the unperturbed metric; the product therefore remains zero to
first order in the perturbation; consequently the LHS must also vanish,
i.e.
$$
     \d  \oint_\X T\cdot\xi  = 0   \eqno(6)
$$
{\it for an arbitrary perturbation $\d g$ and an arbitrary region $\X$}.
But for our $\d g$ this expression itself is by (2) none other than
the difference $E(g)|_{\Sigma_0} - E(g')|_\Sigma$, which accordingly must
vanish.  In other words $E'=E$ or $\d E = 0$, where now $\d E$ just means
the variation in $E$ on $\Sigma$ in going from $g$ to $g'$.  This is our
first main result: the total energy is an extremum for any asymptotically
flat stationary solution to the field equations.

\subsection{Application to a spacetime with internal boundary}

Thus far I have been tacitly assuming that the 3-surface $\Sigma$ is a
complete Riemannian manifold possessing a sole asymptotic region.  When
spacetime has more than one asymptotic region, or more importantly for us,
when it has an {\it internal boundary}, the formula (2) for $E$ must
be applied with care.  In order that it correctly furnish the energy
associated with a given $\infty$, $\xi$ must be a time translation there,
but it must vanish at all the other boundaries (including the actual
internal ones and the ideal ones at infinity).  This rule follows from the
prescription that $E$ represents the change in $S$ which results from a
perturbation that {\it rigidly displaces the entire spacetime relative to
the infinity} in question.  Alternatively, it can be derived by reverting
to the formula (4) for the energy of a ``non-patched'' solution, and
converting (4) to a volume integral via Stokes theorem.

In the case of interest the boundary will be the horizon of a black hole
(or holes).  This surface does not represent a physically real ``edge'' of
spacetime, of course, but rather a boundary we impose on the submanifold we
work with, in order to make effective use of the identity (5).
Being a future horizon, this bounding surface (which I will call $H$) will
be null with its future side facing away from the outer
world.\footnote{$^{5}$}
{It is instructive to examine the reasons why the theorem just proved for
spacetimes without boundary does not furnish useful information when black
holes are present.  In the maximally extended Schwarzschild spacetime, for
example, the theorem does apply, but, since there are two infinities the
extremized energy $E$ is the {\it sum} of the masses seen from the
infinities; and this in turn vanishes since the requirement that $\xi$ be
Killing forces it to point backward in one of the asymptotic regions.  Thus
the theorem is indeed obeyed, but yields only the trivial fact that
$\d(zero) = zero$ !  To make $E$ be the physically relevant energy, we
could eliminate the second infinity via an antipodal identification
(leading to a geon spacetime with spatial topology
${\Reals}P^2\times\Reals^3$), but then one would encounter an inconsistency
in trying to extend $\xi$ inside the horizon as a Killing vector: the
identified spacetime would no longer be stationary in the sense required by
the theorem.  Either way, we fail to gain useful information by trying to
apply the theorem to the manifold as a whole.}

Let us now try to generalize the reasoning of the previous subsection to
a region $\X$ formed as before (with future boundary $\Sigma$ and past
boundary $\Sigma_0$) but with an extra internal boundary $H$ representing
the portion of the horizon between $\Sigma_0$ and $\Sigma$.  In referring
to this setup I will denote the 2-surface $\Sigma\cap H$ by $S$, and the
corresponding, but earlier,  2-surface $\Sigma_0\cap H$ by $O$.  (See
Figure 2.)

\epsfxsize=3.6in
\FigureNumberCaption{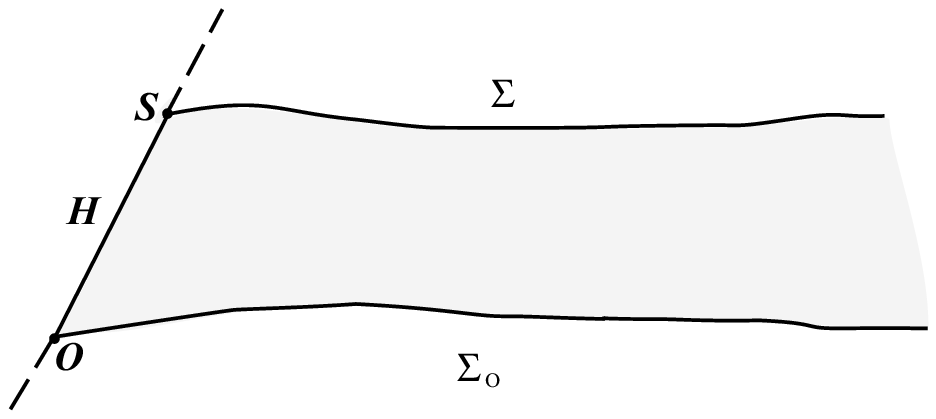}{2}%
{A region $\X$ analogous to that of Figure 1, but truncated at the horizon.
The null surface $H$ is that portion of the horizon between $\Sigma_0$ and
$\Sigma$; its future boundary is the 2-surface $S$ and its past boundary
$O$.}


Now in order to use the identity (6) as we did in the
previous subsection, we need $\xi$ to be a Killing vector of the
unperturbed solution, which contradicts the requirement that it vanish on
$H$ in order that (2) be the total energy.  However $\xi$ {\it is}
a Killing vector at large radii, so there is nothing to stop us from making
it Killing everywhere by use of the relation (3).  Applying this
relation in conjunction with Stokes' Theorem to the region
$\Xi\subseteq\Sigma$ where $\xi$ deviates from being Killing, we
immediately obtain\footnote{$^{6}$}
{by converting the integral of $T\cdot\xi$ over $\Xi$ to an integral
over $\ptl \; \Xi$ of $W\cdot\xi$, then making $\xi$ Killing within $\Xi$,
then converting back to an integral over $\Xi$.}
$$
    - E = \int_\Sigma T\cdot\xi
         + \oint_{S} W\cdot\xi,              \eqno(7)
$$
where now $\xi$ is Killing everywhere and the integral over $S$ appears
because $S$ is the inner boundary of the region $\Xi$.  Expressed in this
manner, the energy appears as a volume integral augmented by a horizon
contribution which it would be natural to desribe as the ``energy of the
black hole''.

Now let us apply to (7) a variation leading from the stationary
solution $g$ to a nearby solution $g'$, and let us temporarily assume that
$g'=g$ in a neighborhood of $S$.  Since the variation of the second
integral in (7) then vanishes trivially, exactly the same proof as
earlier shows that $\d E=0$.  From this it follows immediately that $\d
E$ for a general perturbation {\it can depend only on the value of the
perturbation (and its derivatives) at the horizon itself}, i.e. at the
2-surface $S$.  Notice that  essentially no computation was involved in
reaching this conclusion.

Consider, then, a perturbed solution $g'$ for which $g'-g$ does not
necessarily vanish on the horizon.  We can study its energy by introducing
the same kind of ``interpolating perturbation'' $\d g$ as we used earlier
in the absence of a boundary; however before doing this, it will be
convenient to prepare ourselves by extending the $\Sigma$-integral in
(7) all the way back to $\Sigma_0$ with the aid of the identity
(3).  The result is
$$
   - E = \int_{\Sigma\cup H} T\cdot\xi
       + \int_H G\xi
       + \int_O W\cdot\xi .
                                                              \eqno(8)
$$
Now when we apply the variation $\d$, the first integral in (8) is
unchanged for exactly the same reason as earlier
and we are left with
$$
     - \, \d E =  \d \int_H G \; \xi                  \eqno(9)
$$
(the third integral in (8) being trivially unchanged because $\d g $
vanishes in a neighborhood of $\Sigma_0$).

This is our second main result.  It expresses $\d E$ as the change in the
flux of the fictitious (conserved) energy current $G^a_b\xi^b$ across the
horizon $H$ in going from the stationary to the varied
solution.\footnote{$^{7}$}
{For the case of a rotating black hole, the relevant Killing vector would
be $\xi=t+\Omega\phi$ where $t$, now, denotes the Killing vector which is a
time-translation at infinity, while $\phi$ denotes the rotational one
($\Omega$ being the angular velocity of the horizon).  Hence the variation
$\d E$ in (9) would be replaced by $\d E - \Omega\d J$,
$J$ being the angular momentum.}
Notice that all reference to auxiliary background fields has now dropped
out.

\subsection{Reduction of $\d E$ to an integral on S }

We have already seen on general grounds that $\d E$ must be expressible in
terms of quantities defined only on the 2-surface $S$ in which our
3-surface $\Sigma$ meets the horizon.  To discover the explicit form of
this expression requires us to convert (9) from an
integral over $H$ to one over $S$ alone.  Clearly, it suffices to
re-express it as the integral of a total divergence of some
``potential''.\footnote{$^{8}$}
{Another approach would be to shrink $H$ to a 2-surface by bringing
together the surfaces $\Sigma_0$ and $\Sigma$.}

It turns out that there is a systematic method [8] for
constructing such a potential (and the potential is uniquely determined by
the construction); its applicability is guaranteed by the fact that
$\d(\sqrt{-g}G^a_b\xi^b)$ vanishes for arbitrary variations
$\d{g}$.\footnote{$^{9}$}
{This identity (cf. ref. [7]) is the analog of
eq. (6) for the covariant Action ${1\over{2}}\int{RdV}$,
only expressed in differential form.  It can be derived as such, but it
also follows immediately from eqs.  (6) and (3).}
To apply this method would require only straightforward calculation, but
instead of following this route, we can invoke the Raychaudhuri
equation to evaluate the integral in (9) directly, an
approach which --- though it is less systematic than the method of
reference [8] --- affords a simple explanation of how the
horizon area emerges as the measure of $\d E$.

In essence, all that is involved is using
the Raychaudhuri equation to convert the integrand of (9)
into an expression involving the expansion of the horizon, and then
performing an obvious integration by parts.  In preparation, however, we
need to recall a few definitions and make a convenient choice of gauge in
which to express the perturbation $\d{g}$.

Let us begin by noting that for a non-rotating, stationary black hole
(Schwarz\-schild metric), the timelike Killing vector $\xi$ is automatically
null on the horizon,\footnote{$^{10}$}
{The need for $\xi$ to be null is what forces us to take $\xi=t+\Omega\phi$
in the rotating case, as described in an earlier footnote.}
whence proportional to the null geodesic generators of $H$.  Accordingly,
if we parameterize the latter with an affine parameter $\lambda$, then we
have
$$
         \xi^a = \alpha {dx^a \over d\lambda}  \eqno(10a)
$$
for some function $\alpha$ depending on the choice of normalization for
$\lambda$.  Now although $\alpha$ is not uniquely determined, its
$\lambda$-derivative is, and is given by
$$
       {d\alpha \over d\lambda} = \kappa,  \eqno(10b)
$$
where the black hole's ``surface gravity'' $\kappa$ is defined by the
equation $\xi^b\nabla_b\xi^a=\kappa\xi^a$.

Now in comparing the stationary solution $g$ with the interpolating metric
$g+\d g$, we are free to choose the diffeomorphism-gauge so that the
horizon is the same surface $H$ for both metrics.  In fact we clearly can
arrange that $\xi$ remains a null generator of $H$ with respect to
$g+\d{g}$ and also that $\lambda$ remains an affine parameter along every
such generator.  (For given choices of $\Sigma_0$ and $\d g$, this also
determines to first order in $\d g$ where $\Sigma$ is embedded in the
varied spacetime.)  In this gauge, equations (10a,b) will remain true even
after the variation (with $\kappa$ denoting the surface gravity of the
{\it unvaried} metric $g$, as always.)

Finally we will use the fact that the extensor\footnote{$^{11}$}
{I propose to call ``extensors'', the various tensorial objects which
represent infinitesimal portions of submanifolds in expressions denoting
integrals over such manifolds, for example $d\sigma_a$ and $dV$ in
eqs. (1) and (2), or the codimension-two extensor $dS_{ab}$
which is implicit in the second integral of eq. (7).}
$dS_a$ representing an element of the surface $H$ can be written as
$$
   dS_a = - \dtA \, dx^a \eqno(11)
$$
for a portion of $H$ with cross-sectional area $\dtA$ and extension along
the null direction in $H$ given by the (future pointing) null vector
$dx^a$.

Now we are ready to substitute into (9) the Raychaudhuri
equation for the generators of $H$, namely
$$
    R_{ab}{dx^a\over d\lambda}{dx^b\over d\lambda} =
    - {d\theta\over d\lambda} - {\theta^2\over 2} - {\sigma^2\over 2}.
         \eqno(12)
$$
Since $\theta=\sigma=0$ in the unvaried spacetime ($\theta$ = expansion,
$\sigma$ = shear), only the first term on the RHS of (12) survives
variation; and since $R_{ab}$ vanishes for the unvaried
solution as well, we can write $R_{ab}=\d R_{ab}$ and $\theta=\d\theta$.
Using these facts, we can replace (12) for the metric $g+\d g$ with
the equation
$$
  R_{ab}{dx^a\over d\lambda}{dx^b\over d\lambda} = - {d\theta\over d\lambda}.
                                                \eqno(13)
$$
Noting further that $G^{ab}$ also vanishes for the unvaried metric, we can
now transform (9) as follows:
$$
\eqalignno
{
  - \d E
  &= \d \int_H dS_a G^a_b \xi^b       \cr
  &= \int_H dS_a G^a_b \xi^b         \cr
  &= - \int \dtA (dx)_a G^a_b \xi^b \qquad ({\rm by\ eq.\, (11)}) \cr
  &= - \int \dtA\,(dx)_a G^a_b \alpha {dx^b\over d\lambda}
                                    \qquad ({\rm by\  eq.\, (10a)}) \cr
  &= - \int \dtA\,d\lambda\;\alpha\; G_{ab}{dx^a\over d\lambda}
                                       {dx^b\over d\lambda}            \cr
  &= - \int \dtA\,d\lambda\;\alpha\; R_{ab}{dx^a\over d\lambda}
                                       {dx^b\over d\lambda}            \cr
  &= \int \dtA\,d\lambda\;\alpha\,{d\theta\over d\lambda}
					\qquad ({\rm by\ eq.\,(13)}) \cr
  &= - \int_H \dtA\,d\lambda\; {d\alpha\over d\lambda}\, \theta
     + \int_S \dtA\;\alpha\;\theta                           \cr
  &= - \int_H d\lambda  {d (\dtA)\over d\lambda} \kappa
     + \int_S \dtA \alpha \theta                             \cr
  &= - \int \kappa \int d\lambda  {d (\dtA)\over d\lambda}
     + \int_S \dtA \alpha \theta                             \cr
  &= - \int_S \kappa \d (\dtA)
     + \int_S \dtA \alpha \theta                             \cr
  &= - \kappa \d A + \int_S \dtA \alpha \d\theta             \cr
}
$$
Here in the sixth equality we used that $dx/d\lambda$ is null; in the
ninth we used eq. (10b) and that $\theta$ can be expressed as
$$
  \theta =  {1 \over \dtA} {d(\dtA)\over d\lambda} ;
$$
and in the last equality we used that $\kappa$ is constant on $H$.

Thus we have reduced to $\d E$ to an expression pertaining solely to the
cross section $S$ of the horizon, and this is our third main result:
$$
 - \, \d E = -\kappa \d A + \int_S \dtA\,\alpha\,\d\theta \eqno(14)
$$

\subsection{Locating the horizon}

With equation (14) our work is essentially complete, except
that, in addition to the desired (first) term it contains an integral
depending on $\d\theta$.  In order to realize why this unwanted term is
present, we only have to ask ourselves where we have used the fact that $H$
is actually the horizon of the perturbed solution $g'$, and not just some
random null surface therein.  The point is, of course, that we haven't used
it yet, meaning that the area of $S$ might have changed just because it was
displaced in location without even leaving the unvaried spacetime!  In
order to distinguish such a bogus $\d A$ from the true one, we need a
criterion to locate the horizon with respect to the metric $g'$.  Such a
criterion, I claim, is precisely the requirement that $\d\theta=0$
everywhere on $H$ (where here and henceforth `$\d g$' just means $g'-g$,
not the more complicated interpolating perturbation of earlier
subsections).

In principle this claim, if true, should be derivable from the Einstein
equation, and such a derivation does not look too impractical, at least in
connection with the Schwarzschild metric, whose perturbations are fairly
well understood.  For now however, we will derive $\d\theta=0$ from an
assumption which is a special case of the so-called cosmic censorship
conjecture.  We will {\it assume} that the horizon of the stationary
solution $g$ cannot be destroyed by arbitrarily small perturbations of the
metric.

If this is so (and if it is not, then black holes do not exist in reality
anyway!) then no infinitesimal perturbation of the metric $g$ can make the
expansion $\theta$ negative anywhere, because if it did, then there would
be arbitrarily nearby solutions $g'$ with negative expansion somewhere on
their horizons, but it is well-known that negative expansion implies that
the horizon encounters a singularity in a finite ``time'' (really affine
parameter).\footnote{$^{12}$}
{The argument uses the Raychaudhuri equation: positive convergence implies
infinite convergence in a finite time, implies a generator leaves the
horizon, implies a singularity.}
But if $\theta=\d\theta$ can never be negative, then it can never be
positive either, because a simple change in the sign of $\d g$ will
similarly change the sign of $\d\theta$ (and of course, $-\d g$ will also
be a solution of the linearized Einstein equation).  Hence $\theta$ on the
true horizon must remain zero to first order in any perturbation about a
stationary black hole metric.

\subsection{Summary: the first law}

Our analysis is now complete; let us summarize the highlights.  Using the
identity (6) we first found that $\d E=0$ for any
variation $\d g$ supported away from a cross-section $S$ of the horizon
$H$.  This implied that for general perturbations, $\d{E}$ can depend only
on the behavior of $\d g$ in the neighborhood of $S$.  To evaluate $\d E$
explicitly, we had to re-express the 3-dimensional integral
(9) as the integral of a divergence.  A systematic method
for doing so exists, but we used the Raychaudhuri equation instead, leading
to equation (14).  By invoking the ``stability'' of the horizon
we ``situated'' $H$ within the perturbed spacetime, showing thereby that
the second term in (14) is in fact zero when evaluated on the
correctly identified perturbed horizon.  The remaining term
yields\footnote{$^{13}$}
{For a rotating black hole, the substitution $\xi=t+\Omega\phi$ of earlier
footnotes leads immediately to $-\d E = -\kappa \d A +\Omega \d J$.  For
the charged case, a bit of extra analysis is needed, but again only general
features of the theory are used, without any reference to the explicit form
of the Kerr-Newman metrics (see [3]).}
the so-called ``first law''
$$
           \d E = \kappa \d A.     \eqno(15)
$$

\subsection{Possible further work}

With $A$ identified as entropy, the fact that $\d A=0$ whenever $\d E=0$
can be interpreted as the first-order expression of thermodynamic stability
(in the $\hbar\to 0$ limit), a thermodynamically stable solution being one
which maximizes entropy at fixed energy.  At second order, this
maximization is generically equivalent to
$$
    A'' - \kappa^{-1} E'' \ge 0\ \ {\rm on}\ \ \ker E',  \eqno(16)
$$
where $(\cdot)'$ denotes Fr\'echet derivative.  An interesting problem would
be to try to prove (16) for Schwarzschild (say) by extending the
foregoing analysis to second order.\footnote{$^{14}$}
{In this connection, there might be extra conceptual complications in the
rotating case, associated with the presence of so-called super-radiant
modes.}

Another worthwhile extension of the analysis would be to generalized
gravity theories, including in a Kaluza-Klein setting
(cf. [9]).  There our ``Raychaudhuri trick'' would probably
fail, and one would have to find another trick or fall back on the general
method referred to earlier.  Indeed this general method
[8] merits following up even in ordinary gravity, both as
a ``warmup'' for more complicated Lagrangians, and for the additional
insight it might offer into the origins of the first law itself.

\section{ II. Fractality of the Horizon }

One possible source for the entropy of a black hole is in the fluctuations
of a quantum field propagating near the horizon.  When the field in
question is the linearized metric (``graviton''), the associated entropy is
geometrical in character, but there are many other quantum fields which are
able to contribute as well.  The mechanism in all cases is the same:
fluctuations in the field occur on all scales, and when a fluctuation with
characteristic size $\lambda$ is astride the horizon it sets up a
correlation (``entanglement'') between inside and outside which
metamorphoses into entropy when one ``traces out'' the field modes inside
the black hole in order to obtain the effective density-operator describing
the field outside the black hole [10].

When one tries to compute the value of this entropy for a free field, one
obtains, at first, an infinite result deriving from the fact that free
fields are scale-invariant in the ultraviolet regime, whence an infinite
number of modes contribute with constant entropy per mode.  However, if one
introduces a cutoff at some scale $l$, the entropy takes on the finite
value $S=c A/l^2$, where $A$ is the area of the horizon, and $c$ is a
dimensionless constant of order unity.  Since this gives the right area
dependence, and also the right general magnitude if one chooses
$l=l_{Planck}$, one is tempted to conclude on the one hand that one has
explained black hole entropy, and on the other hand that one has obtained
persuasive evidence for the existence of spatio-temporal discreteness in
nature.

Another thing which speaks in favor of identifying $S$ with some sort of
entanglement entropy is that the prospect of a natural proof of the Second
Law then arises naturally.  Indeed, one can argue that, if full quantum
gravity furnishes us (at some level of coarse-graining) with a
well-defined, autonomously evolving density-operator $\rho$ describing the
outside world, then $-\tr\rho\ln\rho$ necessarily increases as the surface
$\Sigma$ with which it is associated moves forward in time.  The argument
[5] rests on the fact that the total energy is conserved and
determinable from the gravitational field outside the black hole(s), no
matter what may be occuring inside of the horizon (i.e. it rests on
equation (4) or (7) above.  Notice that the entropy does not
change if the codimension-two surface $S$ in which $\Sigma$ meets the horizon
does not move forward along $H$; hence the significance, referred to
earlier, of being able to choose $S$ freely.)

Although the argument just alluded to does not care what degrees of freedom
it deals in (as long as the number of effective external states is finite
at finite total energy), our interest here is in those variables associated
with the fluctuations of quantum fields.  To take seriously their
contribution to the entropy leads to the seeming difficulty that --- for
fixed discreteness scale $l$ --- the magnitude of $S$ would depend on the
total number of fields present in nature, seemingly at odds with the simple
geometrical character of the formula $S=2\pi A$, which just equates the
entropy to the circumference of the unit circle times the area of the
horizon measured in Planck units.  (We take $l_{Planck}=\sqrt{\kappa}$
where $\kappa=8\pi G$ is the ``rationalized gravitational constant'', and
$\hbar=c=1$.)  This simple formula seems more in harmony with a directly
``geometrical'' character for the relevant degrees of freedom, perhaps the
shape of the horizon itself [11], or the configuration of
some underlying discrete structure composing the horizon, such as (the
appropriate portion of) a causal set.

The ``fractal'' picture of the horizon I will describe in a moment grew out
of my wondering whether one could avoid the above ``species dependence
problem'' by somehow writing the quantum fields out of the script in favor
of more suitably geometrical degrees of freedom.  In the meantime it has
become much less clear that there is in fact any difficulty to be avoided,
in view of the observation [12] that a change in the number of
fields would affect not only $S$ but also the renormalized value of
$\kappa\ideq 8\pi G$, and indeed would alter $\kappa$ in just the manner
needed to compensate for the change in the entanglement entropy, leaving
the formula $S=2\pi{}A/\kappa$ still valid.  Although the details of their
argument can be criticized, its overall structure is ``too pretty to be
wrong'', and so is probably correct at some level.  At the same time, it
manifestly ignores the influence of the fluctuations on the horizon itself
(``back reaction''), and to that extent is limited to a semiclassical
regime.

In the picture I am proposing, the number of species is irrelevant for an
entirely different reason, namely for the reason that --- due precisely to
the back-reaction --- the constant $c$ is not constant at all, but rather
depends on the size of the black hole in such a manner as to become
negligibly small for all but Planck sized black holes.  More accurately I
will try to show that the approximation of fixed horizon location and shape
becomes invalid at a length scale much greater than Planckian, namely at a
scale of the magnitude $M^{1/3}$, $M$ being the mass of the black
hole.\footnote{$^{15}$}
{In this and all subsequent formulas, we adopt units such
 that $8\pi{}G\ideq\kappa=1$.}
Below that scale, the field fluctuations become strongly coupled to the
horizon shape, and a semi-classical analysis becomes unreliable.  At the
same time, the shape of the horizon itself becomes ``fractal'' due to the
effects of the fluctuations, perhaps providing the anticipated geometrical
degrees of freedom to ``absorb'' the field ones.

The point is that, at least for free or asymptotically free fields,
fluctuations occur with equal intensity at all sufficiently small scales
$\lambda$.  Given a fluctuation of size $\lambda$, one would expect the
associated energy of magnitude $\sim 1/\lambda$ to induce a concomitant
distortion of the horizon.  Heuristically, we may perhaps picture the
situation as follows.  As one descends in scale, one will reach a
threshold size $\lambda_0$, at which the ``virtual energy'' of a typical
fluctuation will be big enough to distort the horizon shape by an amount
comparable to the size of the fluctuation itself.  Then, like a sleeper who
is uncomfortable in bed and either buries him/herself under the blankets or
pushes them all on the floor, the fluctuation will either pull the horizon
up over itself or (in the case of negative energy-density) drive the
horizon entirely away.  In either case the fluctuation will no longer
overlap the horizon, and it therefore will no longer contribute to the
entanglement entropy.  Moreover, this effect evidently entails a strong
coupling between the horizon shape and the field fluctuations of size
$\lambda\lto\lambda_0$; whence such a fluctuations should not count as
independently ``entangled'' degrees of freedom even if they do happen to
meet the horizon.  We conclude then, that the scale $\lambda_0$ sets a
limit to our understanding of entanglement entropy, and that the only
reliable estimates we can make for the latter pertain to fluctuations with
characteristic sizes greater than $\lambda_0$.

But isn't it obvious that $\lambda_0$ will just turn out to be of Planckian
size in any case?  To begin to answer this question reliably, one would
have to analyze the effect on the horizon of a spacetime ``energy flux
loop'' of characteristic size $\lambda$, situated on or near the horizon
of, say, a Schwarzschild black hole.  Here, I will do something less
accurate but much easier: I will compute for {\it Newtonian} gravity, the
disturbance in the ``horizon'' induced by a small additional mass $m\sim
1/\lambda$ distributed throughout a spatial region of size $\lambda$
located in the vicinity of the horizon.

So let there be present at the origin a spherical mass M, and define its
{\it horizon} as the locus of points where the escape velocity equals unity
(i.e. $c^2$), that is, where
$$
            V = - 1/2,                   \eqno(17)
$$
$V$ being the gravitational potential, $- GM/r$, of the
mass.\footnote{$^{16}$}
{One should presumably conceive of the perturbation as enduring only for a
 time of order $\lambda$, but the associated retardation effects would be
 hard to incorporate in the Newtonian framework, and in any case, they
 would not seem likely to alter the qualitative picture derived from
 treating the horizon as determined by the instantaneous Newtonian
 potential.}
It will be convenient to work, not with $M$, but with the corresponding
``geometrized mass'' or ``Schwarzschild radius'' $R := 2GM$.  In terms of
$R$ we have for a point mass, $V(r)={}-{}R/2r$, so that the horizon occurs
precisely at $r=R$, a well-known coincidence.

Now let us add in the gravitational potential of the fluctuation, to which
for analytical convenience, we will assign the effective mass-density
$\rho=a\lambda/r_1(r_1+\lambda)^3$,
resulting in the potential
$$
                V_1 = {-a/2 \over r_1 + \lambda}.
$$
Here, $a$ is the net geometrized mass of the fluctuation, and $r'$
the distance to its center.  Making the substitution $a=f/\lambda$ ($f$
being some fluctuation-dependent ``fudge factor'' of order unity) yields
finally for  the combined Newtonian potential $V$,
$$
          - 2V = {R\over r} + {f/\lambda \over \lambda + r' }
$$

For simplicity, let us now place the center of the fluctuation where the
unperturbed horizon meets the $y$-axis, and let us also move the origin of
our coordinate system to that point.  Then if we restrict ourselves to the
positive $y$-axis, the potential assumes the particularly simple form
$$
     - 2V =  {R\over y+R} + {f/\lambda \over \lambda + y}.  \eqno(18)
$$
In the approximation that $y,\lambda \ll R$ this reduces to
$$
  - 2V \approx 1 - {y\over R}  + {f/\lambda \over \lambda + y}. \eqno(19)
$$

It is now easy to locate the perturbed horizon by solving the equation
$V=-1/2$ or $-2V=1$.  Working with the approximation (19), we have on
the horizon,
$$
   {y\over\lambda}({y\over\lambda}+1)={fR\over\lambda^3},   \eqno(20)
$$
so that, if we denote by $h=y$ the height of the bulge raised in the
horizon by the fluctuation, we see immediately that the relative height
$h/\lambda$ depends only on the characteristic combination of parameters
$fR/\lambda^3$.  Moreover it is clear that $h/\lambda$ is of order unity
when $fR/\lambda^3$ is, and that it becomes small for $fR/\lambda^3\ll{}1$.
In other words, the threshold scale we are looking for is in fact
$$
                \lambda_0 \sim R^{1/3}
$$
(where I have omitted $f$ since it is in any case of order unity).  The
width of the bulge can be determined similarly.  Indeed, one finds after a
little algebra that the profile of the bulge is determined by the equation,
$$
   {y\over\lambda}
      \left( 1 +
                 \sqrt { ({x\over\lambda})^2 + ({y\over\lambda})^2}
      \right)
      = f {R\over\lambda^3}.
$$
{}From this, one sees that the characteristic parameter $fR/\lambda^3$
governs the shape of the bulge as well as its height, and that the width of
the bulge is comparable to $\lambda$ when $\lambda\lto\lambda_0$.

To summarize: The size and shape of the bulge in the horizon raised by the
fluctuation depends on the ratio $\lambda / \lambda_0$.  For
$\lambda\ll\lambda_0$ the fluctuation raises a bulge much smaller than
itself, whereas for $\lambda\gg\lambda_0$ it is (in our Newtonian picture)
much larger.  In particular, the bulge becomes comparable to the size of
the fluctuation precisely when $\lambda \sim \lambda_0$.  This conclusion
does not depend on the specific profile chosen for the effective mass
density of the fluctuation.  A delta-function would lead to the same
conclusion, as would a dipolar source with vanishing total energy (perhaps
more appropriate as a model of a virtual fluctuation of a quantum field).
And it appears that full general relativity again yields a similar
relationship between scale $\lambda$ and distortion height $h$ if one makes
the drastic approximation of spherical symmetry.

The formula (20), if taken literally, implies that a fluctuation on
scale $\lambda\ll\lambda_0$ induces a distortion of the horizon much
greater than its own size.  However it seems implausible that such an
effect would be present in a fully relativistic setting, where retardation
effects would make such extreme ``action at a distance'' by the fluctuation
appear very unrealistic, and one would not expect the influence of a
fluctuation to extend much beyond its immediate vicinity.  If this is
correct, then it becomes plausible that the actual perturbations in the
horizon due to fluctuations of size $\lambda$ would themselves be of size
$\lambda$ for all $\lambda\lto\lambda_0\sim{}M^{1/3}$.  The resulting
structure of the horizon could then be described as fractal on scales
between $1$ and $M^{1/3}$ (it being doubtful whether spacetime itself
exists as a continuous manifold on scales below unity).  In principle there
is no limit to how large this scale-invariant wrinkling could grow if
sufficiently massive black holes were available, but unfortunately the
prospect of human-sized fluctuations in the horizon disappears when one
plugs in the numbers.  The wrinkles on a solar mass black hole, for
example, would only reach a scale of around $10^{-20}$cm, and for the
fluctuations to attain a size of 1cm, a black hole of the absurd mass of
$10^{91}$ grams would be called for.


\bigskip\noindent
In conclusion, I would like to thank R. Salgado for his indispensable aid in
preparing the figures.  This research was partly supported by NSF grant
PHY-9307570.

\singlespace
\vskip 0.5truein
\centerline           {\bf References}
\par\nobreak
\medskip
\noindent
\parindent=0pt
\parskip=10pt


[1]
  Schutz, B.F. and R.D.~Sorkin,
  ``Variational Aspects of Relativistic  Field Theories, with Application
    to Perfect Fluids'',
   {\it Annals of Phys.} (New York) {\bf 107}:1-43 (1977)

[2]
J.~Bekenstein,
 ``Do we understand black hole entropy?'',
   gr-qc/9409015

[3]
R.D.~Sorkin and M. Varadarajan,
``Energy Extremality in the Presence of a Black Hole'',
  (in preparation)

[4]
D. Sudarsky and R. Wald, Phys. Rev. D {\bf 46}, 1453 (1990)

[5]
 R.D.~Sorkin,
 ``Toward an Explanation of Entropy Increase in the
    Presence of Quantum Black Holes'',
    {\it Phys. Rev. Lett.} {\bf 56}, 1885-1888 (1986);
  see also
   R.D.~Sorkin
 ``Forks in the Road, on the Way to Quantum Gravity'', talk
   given at the conference entitled ``Directions in General Relativity'',
   held at College Park, Maryland, May, 1993,
   (Syracuse University preprint SU-GP-93-12-2).

[6]
 R.D.~Sorkin,
``Conserved Quantities as Action Variations'',
  in Isenberg, J.W., (ed.),
  {\it Mathematics and General Relativity}, pp. 23-37
  (Volume 71 in the AMS's Contemporary Mathematics series)
  (Proceedings of a conference, held June 1986 in Santa Cruz, California)
  (Providence, American Mathematical Society, 1988)

[7]
 R.D.~Sorkin,
 ``The Gravitational-Electromagnetic Noether-Operator and
     the Second-Order Energy Flux'',
  {\it Proceedings of the Royal Society London A} {\bf 435}, 635-644, (1991)

[8]
  R.D.~Sorkin,
  ``On Stress-Energy Tensors'',
    {\it Gen. Rel. Grav.} {\bf 8}:437-449 (1977)

[9]
 Lee, J. and R.D.~Sorkin,
 ``A Derivation of a Bogomol'ny Inequality in Five-dimensional
     Kaluza-Klein Theory'',
   {\it Comm. Math. Phys.} {\bf 116}, 353-364 (1988);
 Bombelli, L., Koul, R.K., Kunstatter, G., Lee, J. and R.D.~Sorkin,
           ``On Energy in Five-dimensional Gravity'',
          {\it Nuc. Phys.} {\bf B289}, 735-756 (1987).

[10]
 R.D.~Sorkin,
``On the Entropy of the Vacuum Outside a Horizon'',
   in B. Bertotti, F. de Felice, Pascolini, A., (eds.),
   {\it General Relativity and Gravitation}, vol. II, 734-736
   (Roma, Consiglio Nazionale Delle Ricerche, 1983);
 Bombelli, L., Koul, R.K., Lee, J. and R.D.~Sorkin,
``A Quantum Source of Entropy for Black Holes'',
  {\it Phys. Rev.} {\bf D34}, 373-383 (1986).

[11]
See the second reference in [10], p. 374.

[12]
L.~Susskind and J.~Uglum, {\it Phys. Rev.} D {\bf 50}:2700 (1994)

\end